\documentclass{svjour3}

\RequirePackage{graphicx}

\usepackage{mathptmx,latexsym,amsfonts}
\usepackage[numbers,comma,square]{natbib}
\newcommand\commentout[1]{}
\newcommand\eqref[1]{(\ref{#1})}

\begin{document}

\title{Spacetime invariants and their uses}

\author{Malcolm A.H. MacCallum} % \thanksref{e1,addr1}}

\institute{School of Mathematical Sciences, Queen Mary University of
London \email{m.a.h.maccallum@qmul.ac.uk}\label{addr1}}

\date{Received: date / Accepted: date}

\maketitle

\begin{abstract}
There are various types of global and local spacetime invariant in
general relativity. Here I focus on the local invariants
obtainable from the curvature tensor and its derivatives. The number
of such invariants at each order of differentiation that are
algebraically independent will be discussed. There is no universally
valid choice of a minimal set. The number in a complete set will also
be discussed. The invariants can then be used to characterize
solutions of the Einstein equations (locally), to test apparently
distinct solutions for equivalence, and to construct solutions. Other
applications concern limits of families of spacetimes, and the
characterization of horizons and singularities. Further uses are
briefly mentioned.
\end{abstract}

\section{Introduction and motivation}
This lecture developed from one given some years ago at the retirement
party of my friend, colleague and co-author Dr.\ Eduard (``Eddie'')
Herlt. At the Lahore meeting, time did not permit a complete
exposition of all aspects of the subject. This text follows what was
said rather than what might have been said in an expanded version. I
intend to write up the expanded version as a review article in due
course.

I could have been more precise in my title, at the cost of being
rather lengthy. There are many occurrences of the word 'invariant' in our
field, e.g.:\\
gauge-invariant (in gauge theory)\\
gauge-invariant (in perturbation theory)\\
invariant under a symmetry\\
Lorentz-invariant\\
scale-invariant.\\
And I could be talking about some global conserved quantity, without a
well-defined local density, e.g.\ Bondi mass.

What I actually meant was local, geometric, invariants of spacetime:
essentially curvature invariants.

I started to be seriously interested in this area when we were writing
the first edition of the exact solutions book \cite{KraSteMac80} in
the late 1970s.  One of the big problems we faced was that of
identifying the same solution when found with different assumptions or
for different reasons, and presented in different coordinate
systems. I realised at the time that the work of the Stockholm group
on invariant classification, which I will mention later, and which I
first heard about in the late 70s, held the key.

This first application, as far as I was concerned, the ``equivalence
problem'', was what got me started. 

\section{Mathematics}

\subsection{Definitions of invariants}

Christoffel proved in 1869 that scalars constructed from the metric and its
derivatives must be functions of the metric itself and the Riemann
tensor and its covariant derivatives.

The first examples to spring to mind are {\bf scalar polynomial (s.p.)
  invariants}, 
such as $R_{ab}R^{ab}$ or $C_{abcd}C^{cdef}{C_{ef}}^{ab}$. Often
when people just say ``invariant'' they mean ``s.p. invariant''.

However, these are not adequate in all circumstances, as one can see by
noting that $pp$ waves and flat space both have all
scalar polynomial invariants, of all orders, equal to zero \cite{JorEhlKun60}.

Fortunately they are not the only choice.
An important alternative is to use ideas due to Cartan, as follows.
As a side benefit, in my view an important one, Cartan invariants
require less calculation, in general.

Let $F({\cal M})$ denote a ``suitable'' frame bundle over a spacetime ${\cal M}$
(i.e.\ take the set of all frames at each point) and ${\cal R}^q$ be the set
$\{R_{abcd},\,R_{abcd;f} ,\,\ldots,R_{abcd;f_1f_2\cdots f_q}\}$ of the
components of the Riemann tensor and its derivatives up to the $q$th
in a frame.

Choose from $F({\cal M})$ in a canonical and invariant way,
e.g.\ use the principal null directions of the Weyl tensor, when they
are distinct, as the
basis vectors. The
resulting 
\[\{R_{abcd},\,R_{abcd;f} ,\,\ldots,R_{abcd;f_1f_2\cdots
  f_q}\}\] are called the {\bf Cartan invariants}. They are scalars,
because the frames are invariantly defined, e.g.\ if {\bf a}, {\bf b},
{\bf c} and {\bf d} are basis vectors (not necessarily all distinct)
of the chosen frame, one of the Cartan invariants is
$R_{ijkl}a^ib^jc^kd^l$ .  The idea is like characterizing a symmetric
bilinear map (matrix) by its eigenvalues.

\subsection{Independence of (s.p.) invariants}
In a manifold ${\cal M}$ of $n$ dimensions, at most $n$ scalar invariants
can be {\bf functionally independent}, i.e.\ independent functions on
${\cal M}$.

The number of {\bf algebraically independent} scalar polynomial
invariants, i.e.\ s.p.\ invariants not satisfying any polynomial relation
(called a {\em syzygy}), is rather larger, as we shall see
next. (A set of algebraically independent Cartan invariants in a
general spacetime, written using the Newman-Penrose spinor formalism, is
given in \cite{MacAma86}: it takes fully into account the Ricci and
Bianchi identities.)

Larger still is the size of a complete set of s.p.\ invariants (a
finite Hilbert basis): such a set $\{I_1,\,I_2,\ldots,\,I_k\}$ is {\bf
  complete} if any other such s.p.\ invariant can be written as a {\em
  polynomial} in the $I_j$ but no invariant in the set can be so
expressed in terms of the others.

One way to find the number of algebraically independent scalar
invariants is to consider Taylor expansions of the metric and
of the possible coordinate transformations \cite{Sik76}; another
way follows Hilbert \cite{Sik06}.

The result \cite{Tho34}
is that in a general $V_n$ the number of algebraically
independent scalars constructible from the metric and its derivatives
up to the $p$th order (the Riemann tensor and its derivatives up to
the $(p-2)$th) is 0 for $p=0$ or $p=1$ and
\[
N(n,p) = \frac{n[n+1][(n+p)!]}{2n!\,p!} - \frac{(n+p+1)!}{(n-1)!\,(p+1)!}
+ n,
\]
for $p \geq 2$, except for $N(2, 2) = 1$.  Thus in a general
space-time the Riemann tensor has $N(4,2)=14$ algebraically
independent scalar invariants. In
particular cases the number is
reduced.

The origin of many syzygies can be understood in terms of the
vanishing of any object skewed over $(n+1)$ indices in $n$ dimensions
\cite{Har95,Bon98,Edg99}. In a series of papers
\cite{LimCar04,CarLim06,LimCar07} Carminati and Lim
have given a detailed discussion
of the syzygies for scalar polynomial (s.p.)~invariants of the Riemann
tensor, using graph-theoretic techniques.

A given invariant may be written in more than one way, due to
symmetries, and other relations between components, of the Riemann
tensor and its derivatives. This is essentially a problem in
representations of the permutation group. The issue is to select a
canonical representative of each equivalence class in the orbit under
permutations.

One wants to select a canonical set of invariants and then express any
other invariant in terms of canonized members of the algebraically
independent set.

Several methods have been used to do this for s.p.\ invariants,
e.g.\ by H\"ornfeldt, by Fulling et al., by Ilyin and Kryukov, and by
Dresse.  The method most readily available is due to Portugal \cite{Por98}. It
has been implemented in Maple${}^{\rm TM}$ and {\it
  Mathematica}${}^{\circledR}$ by Martin-Garcia {\em et al,} and is
distributed in xAct, a package for use with {\it
  Mathematica}${}^{\circledR}$. It has (e.g.) been applied to all
objects with up to 12 derivatives of the metric \cite{MarYllPor08}.

Any complete set of s.p.\ invariants of the Riemann tensor, and any
set which {\em always} contains a maximal set of independent scalars,
contains redundant elements.  Hence all the old papers giving specific
sets of 14, in 4 dimensions, are inadequate.

The smallest known set of s.p.\ invariants for a general Riemann
tensor that always contains a maximal set of algebraically independent
scalars consists of 17 polynomials \cite{ZakMcI97}, though 16 suffice
for perfect fluids and Einstein-Maxwell fields. There is a special
subclass of spacetimes \cite{ZakCar01} which require 18.

The smallest set of s.p.\ invariants known to be complete contains 38
scalars \cite{Sne99}, and Lim and Carminati \cite{LimCar07} proved that it is
minimal.

\subsection{Computation of invariants}
The s.p.\ invariants are expressions whose contractions hide very
large numbers of individual terms, and are therefore hard to
calculate. It is thus very useful to adopt a method that reduces
the number of terms. One can use a bivector formalism
(see e.g.\ \cite{ZakMcI97}).  Another way, in 4 dimensions, is to use the
Newman-Penrose complex spinor (or null tetrad) formalism, the GHP formalism, or
related ideas: this has analogues in other dimensions (e.g.\ in three
dimensions one can use real two-component spinors: see \cite{SouFonRom08}).

For higher dimensions it may be efficient to use parallel computing,
i.e.\ to split the calculation into subproblems e.g.\ by fixing some
of the dummy indices, and then sending the subproblems to separate
processors \cite{Koe06}.

A simple practical way in many cases is to use Cartan invariants
instead. They are quite cheap to calculate {\bf provided} the frame choice
is easy to calculate with, in particular because one never has to multiply
curvature components, or their derivatives, together.  In a general
case, either s.p.\ or Cartan invariants would characterize the
spacetime, as we shall see below. so the choice is really a question
of convenience or purpose.  There is an open research problem
about which is actually more computationally efficient (or for which cases).

In four dimensions we can always choose a frame, for example by the
principal null directions of the Weyl tensor, and thus compute Cartan
invariants, but in general this may entail using the unpleasant
algebraic expressions which arise from the general formulae for
solutions of quartics. However, in practice many solutions give way easily. Note
that in 5 dimensions or more there is no guarantee we can calculate
frames of a specified type even if we can show they must exist, as
there is no general algebraic formula for solutions of quintics.

\subsection{Do scalar polynomial invariants suffice?}

Until 2009, I would have said definitely not. 
I mentioned earlier that $pp$ waves and flat space both have all
scalar polynomial invariants, of all orders, equal to zero. In fact
all vacuum type N and III metrics with $\rho=0$ have this property
\cite{PraBi01}.
% See Nita and Robinson for a way to classify using this in reverse
% Indeed one can find all such cases (see later).
There are also metrics which have equal non-zero
s.p.\ invariants e.g.\ \cite{Sik96,Pra99,Her04}.
%Schmidt 2001

 These ambiguities 
are associated with the indefiniteness of the metric and the
non-compactness of the Lorentz group \cite{Sch98}.

Coley {\em et al.} \cite{ColHerPel09} gave an argument that
spacetimes are completely characterized by their s.p.\ invariants,
except for spacetimes in the Kundt class.  Although their proof seems
to me to have a gap, and I have not yet been able to see if it has been
filled, the result seems correct.

In particular Coley and collaborators have followed up by a
substantial number of papers, which I did not have time to review
comprehensively in my Lahore talk. These consider not only 4D
spacetimes, but higher dimensional spaces and various signatures.  I
just mentioned a few to give the flavour. In the initial paper
\cite{ColHerPel09}, they also discuss what properties a given set of
invariants characterize.

\begin{itemize}
\item Hervik and Coley \cite{HerCol10}
%Hervik and Coley (2010)
 show that any metric with an
  analytic continuation to the pure Riemannian case is completely
  characterized by its s.p. invariants%
\item Coley {\em et al.} \cite{ColHerDur11}
% Coley et al (2011)
 give a detailed analysis of the generic 5D
  spacetimes.%
% 1011.2175 is not used here
\item Hervik \cite{Her11}
%Hervik (2011)
 gave arguments that spacetimes not characterized by
  s.p. invariants are limits of families that are so characterized.
% 1107.3210
\end{itemize}

It may be worth noting that since it is obvious that the algebraically
independent set of Cartan invariants given in \cite{MacAma86}
completely and uniquely specify the s.p.\ invariants, the result of
\cite{ColHerPel09} amounts to saying that in general the converse is
true, i.e.\ that the s.p.\ invariants uniquely determine the Cartan
invariants (given the procedure for the choice of frame).

The work just described picked out the Kundt class as the
exceptions. These are spacetimes which have a null vector field which
is geodesic, and expansion-, shear- and twist- free.  In a paper
\cite{ColHerPap09} which appeared just before \cite{ColHerPel09} this
class was discussed in detail. For it, one has to use the Cartan
approach or some equivalent.

In \cite{ColHerPap09} Coley {\em et al.} discussed the Petrov D
examples in the Kundt class.  Podolsky and Svarc \cite{PodSva13} gave
a classification of Kundt metrics in arbitrary dimension, and later
discussed their physical interpretation. A subclass of ``universal
spacetimes'' was discussed by Hervik {\em et al.} \cite{HerPraPra14}:
universal in that they solve vacuum equations of all gravitational
theories with Lagrangian constructed from the metric, the Riemann
tensor and its derivatives of arbitrary order. Hervik {\em et al.}
\cite{HerHaaYam14} found all cases where there exists a continuous
family of metrics having identical [scalar] polynomial invariants.
% 1303.0215 and 1306.6554 and 1311.0234 and 1410.4347
% Not mentioned: 1109.2551, 1210.0746
%
% [Note: I have yet to understand all the details of these lengthy papers.]

\section{Uses of spacetime invariants}
\subsection{The Equivalence Problem}

Now let me return to my original motivation, i.e.\ identifying the same
solution when found with different motivations and assumptions and
presented in different coordinate systems. It could be
considered either a mathematical or a physical problem. It is essentially
resolved by the following theorem (in terms of Cartan invariants).

\begin{theorem} {\rm (Sternberg \cite{Ste64}, Ehlers \cite{Ehl81})} Given
  two spacetimes, 
  ${\cal M}$ and $\overline{{\cal
  M}}$, each expressed using some frames $E$ and $\overline{E}$, then
  there is an isometry which maps $(x,\,E)$ to
  $(\overline{x\vphantom{E}},\,\overline{E})$ if and only if ${\cal
  R}^{p+1}$ for ${\cal M}$ is such that the independent quantities and
  the functional relations are the same for ${\cal R}^{p+1}$ and
  $\overline{{\cal R}}^{p+1}$, where $p$ is the last derivative at
  which a new functionally independent entry in ${\cal R}^{p}$ arises.
\end{theorem}

See \cite{SteKraMac03} for a more careful statement.

One could therefore calculate everything on a suitable frame bundle.
\begin{itemize}
\item Christoffel used coordinate frames. This works but the dimension of
$F({\cal M})$ is large.
\item Cartan proposed using frames with constant
scalar products, e.g.\ null frames or orthonormal frames, which is
better (hence the credit for Cartan invariants).
\item In 1965 Brans \cite{Bra65}
proposed practical use of this
method via lexicographic indexing of components.
\item Later Brans and, more fully, Karlhede \cite{Bra77,Kar80a}
  realised that a more efficient way to implement the idea was to
  restrict the frames wherever possible by only allowing canonical
  forms of the curvature and its derivatives (i.e.\ go to Cartan
  invariants).
\item This makes the
  whole thing manageable (with a computer) and it was first
  implemented by {\AA}man in 1979 \cite{KarAma79}.
% (Karlhede and {\AA}man 1979).
\end{itemize}

Given the Coley et al. results, we now know we could in general work
only with s.p.\ invariants, but the present software \cite{MacSke94,PolSkeDin00}
uses the Cartan method.  In an algebraically general spacetime, one
can choose the principal null directions of the Weyl tensor to fix the
frame, or in a conformally flat spacetime one can begin with
eigenvectors of the Ricci tensor. In the actual implementaion we use
the Weyl PNDs, where they exist, as the first choice.

Many details have been considered (by {\AA}man, myself, Skea, d'Inverno,
McLenaghan, Pollney and others) in specifying frames and in making
the computations more efficient. The main aspects are the enumeration
of canonical forms, the tests of which canonical form applies, and the
transformation of the non-canonical to the canonical. The literature
cited in the exact solutions book gives more information.

A first extension is to maps more general than isometries,
e.g.\ homotheties \cite{Kou92} or conformal equivalence \cite{Lan93}. One can of course tackle Euclidean metrics and metrics
in any dimension, though the details of making the method work
efficiently are different (e.g.\ \cite{Kar86}).

The above ideas are in fact a special case of Cartan's general
procedures (see e.g.\ \cite{Car46}), which apply to other situations
which can be expressed in differential forms and a connection
or similar structures. In particular it applies to the equivalence of
(systems of) differential equations, under coordinate transformations
(see the books of Gardner \cite{Gar89} and Olver \cite{Olv95}).

Another context is that of gauge theories of physics, the gauge
potential being just a connection \cite{Kar86a}.

One interesting issue for spacetimes is how large $p$ in the theorem
has to be. Karlhede \cite{Kar80a} showed $p\leq 7$. For a long time examples
suggested $p \leq 3$. At the time of the second edition of the exact
solutions book \cite{SteKraMac03} an example with $p=5$ was known, as
were a number of bounds for subclasses. We quoted results suggesting
that $p\leq 5$ was generally true.

In 2009 (the result was actually first announced in 2007), 
Milsom and Pelavas % (2009) 0711.3851
\cite{MilPel09} showed that
the original bound of 7 is sharp, by giving a (unique) example. This is still
small enough for calculation to be practical. The example is of course in
Kundt's class, as are those further examples they found where $p=6$.

As well as giving a way to compare two spacetimes, the theorem above
also implies that the Cartan invariants uniquely classify and
determine the local geometry. This implies that all local properties
are encoded in this information.  I will rather briefly discuss some
of the possible resulting applications. Aspects I was not able to
cover in Lahore included: applications to three-dimensional spaces,
incuding the use in checking junction conditions; applications in
numerical relativity; applications to cosmology and to perturbations;
use in constructing alternative theories of gravity; and applications
to internal state spaces of thermodynamics.  I think there must be
more we have not yet explored.

Note that global topology is of course not determined by the
invariants we have been considering (local
invariants cannot distinguish a plane and a flat torus) and nor are
continuations which do not respect the required differentiability for
the theorem above. 

My original motivation can now be carried out for many cases.  I used
these methods, for example, to disentangle the known exact solutions
for inhomogeneous cosmologies with a symmetry group $G_2$. It would be
good to have a complete online catalogue of (at least) the solutions
in \cite{SteKraMac03}. Skea and Lake have
independently implemented systems to facilitate this.  Properties such
as the isometry group can also be found
\cite{KarMac82,AraDraSke92,BonCol92}: for examples see
e.g.\ \cite{Sei92,MacSik92}. Similar results apply for homotheties
\cite{Kou92}.

\subsection{Finding solutions}
The same ideas can be used to {\em find} solutions, by working out
consistent sets of Cartan invariants and then integrating to get a
metric. Examples are mentioned in the exact solutions book.  Machado
Ramos and Edgar \cite{MacEdg05,EdgMac07} used these ideas, implemented
in their invariant operator formalism, to find pure radiation
solutions of types N and O with a cosmological constant.

Coley and collaborators looked for all spacetimes in which all
s.p.\ invariants vanish (which they called VSI spacetimes although not
all Cartan scalar invariants vanish). %cf Page
In the 4-d Lorentzian case
these give all spacetimes indistinguishable from flat space by
s.p.\ invariants \cite{PraPraCol02}. The higher dimensional cases
give examples  in supergravity and string theory.
A closely related series of papers has studied the spacetimes with
constant s.p.\ invariants (CSI spacetimes). For a review of this work
see \cite{Col08}. 

\subsection{Limits of families of spacetimes}
Another aspect, described first by Paiva {\em et al.}
\cite{PaiRebMac93}, is the use of invariants to work out the limits of
families described by parameters. This enables one to find all
possible limits in a coordinate-free way, whereas previous treatments
were in effect trial and error. This is another area where further
work would be useful.

I did not give more detail in my Lahore lecture, but passed quickly on
to limiting points within spacetimes.

\subsection{Black hole and other horizons}
Karlhede {\em et al} \cite{KarLinAma82} first noted that $R_{abcd;e}R^{abcd;e}=0$ at
the Schwarzschild horizon (so a prudent space traveller might monitor
that). Skea in his thesis \cite{Ske86} noted that this is not true for other
horizons (a point rediscussed by Saa \cite{Saa07} for higher-dimensional
static cases: Saa also found points where $R_{abcd;e}R^{abcd;e}=0$ but
which are not horizons).
% See extra refs in Saa
 Lake \cite{Lak03} continued the work on Kerr by considering first derivative
invariants, and found their vanishing characterized the horizons.

Related to my arguments against the claims of Antoci and others that
the Schwarzschild horizon was singular, I proposed in 2006
\cite{Mac06} a new
test for occurrence of a horizon using ratios of Cartan invariants
which works in all cases of Petrov type D. I believe it works more
generally, but this needs further study. Maybe it could also be useful
to numerical relativists (and space travellers).

More recently Abdelqader and Lake \cite{AbdLak13,AbdLak14}
have given an invariant characterization of the Kerr horizon, and,
prompted by that, Page and Shoom \cite{PagSho15} have given another for
stationary black holes. I have not yet had the opportunity to check how
these relate to earlier characterizations.

Moffat and Toth \cite{MofTot14} considered the relation of the ``Karlhede
invariant'' (i.e.\  $R_{abcd;e}R^{abcd;e}$) to discussions of a
``firewall'' at the horizon.

\subsection{Singularities}
Singularities in general relativity are defined to occur when a causal
geodesic cannot be continued to infinite affine parameter values even
when the spacetime is maximally extended -- ``geodesic
incompleteness'' (for the reasons for this definition see \cite{Ger68}).

It is well-known that this happens if an s.p.\ invariant of the
Riemann tensor itself (not its derivatives) blows up along the
geodesic. This does not however mean that:\\ (i) an ``infinite''
Riemann tensor implies a singularity, or\\ (ii) if an invariant blows
up, there is a singularity, or\\ (iii) at all singularities, an
s.p.\ invariant of the Riemann tensor blows up.

\begin{enumerate}
\item An ``infinite'' Riemann tensor does not imply a
singularity. Geodesics can be continued across a delta function
curvature modelling a thin shell or an impulsive gravitational wave.
\item The blowing up of an invariant does not imply a singularity. For
example \cite{Mac06} the invariant $1/R_{abcd;e}R^{abcd;e}$ blows
up at the Schwarzschild horizon, but the horizon is not singular.
\item There are singularities at which no s.p.\ invariant of the
  Riemann tensor blows up. At ``whimper'' singularities an invariant
  involving first derivatives does, though \cite{Sik78}. An example
  was studied by Podolsky and Belan \cite{PodBel04}.
\end{enumerate}

Question: when does blow up of higher derivative invariants imply a
singularity? 

Another application is to ``directional'' singularities, where
a singular point apparently has directionally dependent
limits. Scott and Szekeres \cite{ScoSze86,ScoSze86a} showed 
that the
directional singularity of the Curzon metric hid more extended
regions at whose boundary the original coordinates broke down. My
student Taylor \cite{Tay05} showed that such cases could be
appropriately ``unravelled'' by using level surfaces of Cartan
invariants to define new coordinates.
Lake \cite{Lak03} used the Weyl tensor s.p.\ invariants to show that the Kerr
singularity was not directional.
% Cherubini et al (2002) discuss
% and interpret the same invariants.
%\end{frame}

With Coley and others \cite{ColHerLim09}, I
considered ``kinematic singularities'' (where fluid expansion blows
up),
giving examples
in which, given an integer $p$, the Cartan (and hence s.p.) scalars can
be finite up to the $p$-th derivative, but not the $(p+1)$-th.

Geodesic continuation needs a $C^{2-}$ metric. In
invariantly-defined frames the connection coefficients (which would be
$C^{1-}$) are typically
expressible as ratios of first derivative Cartan invariants to zeroth
derivative ones. We know that there are ``intermediate'' or
``whimper'' singularities where s.p.\ invariants of the Riemann tensor
do not blow up, while %in the explicit examples
   s.p.\ invariants of
the first derivatives of the Riemann tensor do. Hence I conjecture
that under some suitable differentiability conditions:

{\bf Conjecture:} Spacetime singularities are either locally
extendible or at least one Cartan invariant in ${\cal R}^1$ has an
infinite limit along any curve approaching the singularity.

%\bibliographystyle{spmpsci}      % mathematics and physical sciences, numeric
% \bibliographystyle{spphys}       % APS-like style for physics, numeric
% \bibliography{GRstrings,cartan,lahore}   % name your BibTeX data base

\end{document}